\documentclass[aps,prl,twocolumn,showpacs]{revtex4}
\usepackage{epsfig, amsmath, amsfonts, amssymb, bm, graphicx}

\begin{document}

\DeclareGraphicsExtensions{.eps,.EPS,.jpg,.pdf}
%%%%%%%%%%%%%%%%%%%%%%%%%%%%%%%%%%%%%%%%%%%%%%%%%%%%%%%%%%%%%%%%%%%%%%%%%%%%%%%%%%%%%%%%%%%%%%%%%%%%%%%%%%%%%%%%%%%%%%%%%%%%%%%%%%%%%%%%%%%%%%%%%%%%%%%%%%%%%%%%%%%%%%%%%%%%%%%%%%%%%%%%%%%%%%%%%%%%%%%%%%%%%%%%%%%%%%%%%%%%%%%%%%%%%%%%%%%%%%%%%

\title{Collective spin modes of a trapped quantum ferrofluid}
\author{S. Lepoutre, L. Gabardos, K. Kechadi, P. Pedri, O. Gorceix, E. Mar\'echal, L. Vernac, B. Laburthe-Tolra}
\affiliation{1 Universit\'e Paris 13, Sorbonne Paris Cit\'e, Laboratoire de Physique des Lasers, F-93430
Villetaneuse, France and 2 CNRS, UMR 7538, LPL, F-93430 Villetaneuse, France}

%\pacs{03.75.Mn , 05.30.Jp, 67.85.-d, 05.70.Ln}
\date{\today}

\begin{abstract}
We report on the observation of a collective spin mode in a spinor Bose-Einstein condensate. Initially, all spins point perpendicular to the external magnetic field. The lowest energy mode consists in a sinusoidal oscillation of the local spin around its original axis, with an oscillation amplitude that linearly depends on the spatial coordinates. The frequency of the oscillation is set by the zero-point kinetic energy of the BEC. The observations are in excellent agreement with hydrodynamic equations. The observed spin mode has a universal character, independent of the atomic spin and spin-dependent contact interactions.
\end{abstract}

\maketitle

Collective excitations are a unique tool for exploring the  collective effects that arise in dilute quantum gases due to interactions between the atoms. One prominent example is superfluidity in Bose or Fermi gases, which can be probed by studying scissors modes \cite{marago2000, derossi2016} or the nucleation of topologically stable vortices \cite{matthews1999,anderson2000,madison2000,abo2001,hodby2002,raman2001,zwierlein2005}. Besides superfluidity, quantum gases generally behave like fluids, and are efficiently described by collision-less superfluid hydrodynamic equations \cite{stringari}.

\begin{figure*}[t]
\centering
\includegraphics[width= 14 cm]{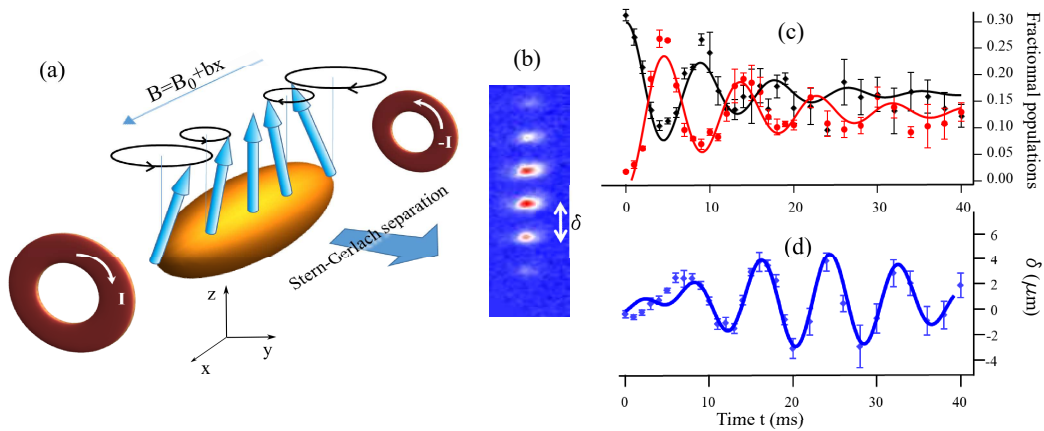}
\caption{(a) A BEC is initially polarized perpendicular to the external magnetic field. Spin collective modes are excited by a magnetic field inhomogeneities (generated by several coils, two of them being sketched). In the rotating frame, these modes correspond to spins oscillating \textit{around their initial direction} (and not around the inhomogeneous field) with an amplitude which linearly depends (in this Figure) on the position $x$. The mode is analyzed after separation of the spin components by a Stern-Gerlach procedure (which produces absorption images such as the one shown in (b)). From the absorption pictures, we extract (c) the dynamics of the population in the different Zeeman states $p_{m_s}$ (for clarity, only $m_s=0$ and $m_s=-3$ are shown), as well as (d) the separation $\delta$ between the Zeeman states, which is associated to the \textit{in situ} spin flux. Solid lines are fits by a damped sinusoid (top), and fits by a sum of two sinusoids (bottom). Here $g \mu_B b/h=30.1$ MHz/m.}
\label{principe}
\end{figure*}

Quantum gases with an internal degree of freedom are new types of quantum fluids \cite{stamperkurn} which, similarly to the cases of superfluid $^3$He \cite{leggett}, exciton-polariton condensates \cite{lagoudakis2009} or  triplet superconductors \cite{jang2011}, can display novel properties due to the enlarged degrees of freedom and the new possibilities for symmetry breaking \cite{kawaguchi2012}. The study of collective excitations of spinor quantum gases is thus expected to display a very rich phenomenology, e.g. the possibility of non abelian excitations \cite{kobayashi}. However, experimental research only recently began, with the notable observations of topologically stable 2D skyrmions \cite{choi2012}, magnons \cite{marti2014}, and half-quantum vortices \cite{seo2015}. Here, we consider an harmonically trapped spinor gas prepared in a non-equilibrium ferromagnetic state with all spins aligned perpendicular to the direction of the external magnetic field; we investigate a collective mode coupling the spin and the orbital degrees of freedom, which is triggered by simply applying a magnetic field gradient.

In addition to the gradient, the experiment is performed in a $\approx$1 G magnetic field, which leads to fast Larmor precession. The outcome of the experiment is best understood in the frame rotating at the Larmor frequency ($\approx$MHz). Using hydrodynamics equations \cite{kawaguchi2010}, we demonstrate that, in the rotating frame, rather than merely undergoing single-particle inhomogeneous precession around the magnetic field, the atoms collectively respond to the gradient in such a way that the local spin oscillates around its original direction with a spatially dependent amplitude. A \textit{trapped magnon mode} is thus spontaneously generated (see Fig. \ref{principe}(a)). The lowest energy mode, for which the amplitude of the local spin oscillation is linear with respect to the position, is characterized by a frequency of order $\hbar/M R^2$ (where $\hbar$ is the reduced Planck constant, $M$ the atomic mass and $R$ the extension of the cloud). This frequency is well below the trapping frequencies of the 3D harmonic potential, a unique feature which never occurs for the lowest energy modes of scalar ground-state Bose-Einstein Condensates (BECs) \cite{stringari} (see however the observed low frequency Tkachenko oscillations in metastable vortex lattices for rotating superfluids \cite{coddington2003}). The collective modes, which are derived from the ferrofluid hydrodynamic theory and found to be in excellent agreement with the experimental data, are universal in that their frequencies depend neither on spin-dependent interactions nor on the atomic spin. Our work thus also illustrates that the hydrodynamic approach provides an efficient unified description of locally polarized spinor gases as quantum ferrofluids.

The experimental setup has been described in \cite{lepoutre}. We start from a BEC comprising $4.10^4$ $S=3$ chromium atoms in the lowest energy spin state $m_s=-3$. We tilt the spin of the atoms orthogonal to the magnetic field $\vec{B}_0$ using a radio-frequency (rf) pulse. This is performed in presence of a magnetic field gradient $b$. However, the Rabi frequency of the rf field is much larger than the field inhomogeneities, so that the rotation is homogeneous across the sample. To characterize the state of the BEC at a given time $t$ after the rf pulse, we then turn off the atomic trap, and perform a Stern-Gerlach separation of the different Zeeman components by applying a magnetic field gradient (which is much stronger than $b$) during the $t_{tof} \approx 7$ ms time-of-flight. Finally, we take an absorption image (Fig. \ref{principe}(b)). On the images, the spatial separation between two consecutive Zeeman components is independent on $m_s$, and thus characterized by a single distance $\delta$.

As shown in Fig. \ref{principe}(c-d), we observe an evolution as a function of $t$ of both the different Zeeman relative populations $p_{m_s}$ and the spatial separation $\delta$. The spin dynamics revealed by the evolution of $p_{m_s}$ is inherently driven by spin-dependent contact and dipolar interactions. On the other hand, $\delta$ primarily depends on the magnetic field gradient during the time-of-flight; the (small) time-dependance in $\delta$ necessarily results from an \textit{in situ} mechanical motion  of the different Zeeman components within the trap (i.e. before Stern-Gerlach separation). As we demonstrate in this paper, the dynamical evolution of the spin population is coupled to such a mechanical motion, which is a consequence of the spin flux generated by the presence of a magnetic field gradient.

\begin{figure}[t]
\centering
\includegraphics[width= 8.5 cm]{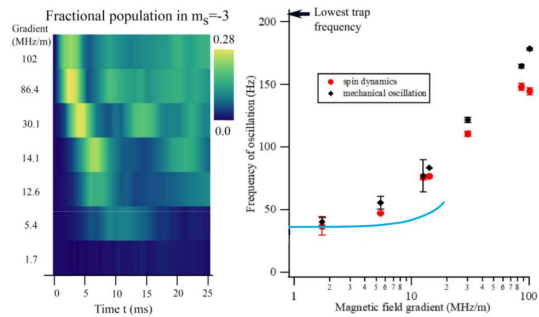}
\caption{Experimental data as a function of the magnetic field gradient. Left: fractional populations in $m_s=-3$ as a function of time, for various magnetic field gradients $b$, showing oscillations for each $b$.  Right: Fitted oscillation frequency associated to population dynamics and spin motion. The solid line results from solving Eq. (\ref{eqnhydro}) beyond the small oscillation approximation \cite{suppmat}, and converges to the result of eq.(\ref{freq}) when $b\rightarrow 0$.}
\label{gradient}
\end{figure}

We investigate the coupled behavior of spin dynamics and orbital motion within the trap by varying the magnetic field gradient $b$. Although complex oscillatory behaviors are obtained when $b$ is large (Fig. \ref{principe} (c-d)), at low gradients we observe a rather simple damped oscillatory behavior for both the population dynamics and the separation  $\delta(t)$, with respective angular frequencies $\omega_s$ and $\omega_m$. We find that $\omega_m \approx \omega_s$ (see Fig. \ref{gradient}), and that these frequencies strongly depend on the magnetic field gradient $b$. The amplitude of oscillation also depends on $b$, and vanishes for $b \rightarrow 0$.

These observations indicate that the interaction with magnetic field gradients has excited a collective mode which couples the spin degrees of freedom (characterized by $p_{m_s} (t)$) to the spatial degrees of freedom (characterized by $\delta(t)$). The behavior at the smallest magnetic field gradients is most intriguing, as we observe that the motion of the different Zeeman states within the trap occurs at a natural frequency well below the trapping frequencies of the harmonic potential. This is a strong indication that the mode is driven by collective effects.

To account for the observed mode, we developed the following theoretical model. Based on the observations in \cite{lepoutre}, we assume that the Bose-Einstein condensate remains locally polarized (ferromagnetic) at all times after tilting the spins, an emergent collective effect which arises due to an energy gap provided by spin-dependent contact interactions. Consequently, the local spin length per atom remains close to $S=3$ at all times, and only the orientation of the spin can evolve as a function of space and time. The quantum gas then behaves like a quantum ferrofluid, which can be efficiently described by the hydrodynamic equations of a ferromagnet \cite{kawaguchi2010}. We further assume that the global density $n_{tot}$ remains constant in time (in absence of losses), a property which is typically displayed by ferrofluids, and which our numerical simulations of the 3D Gross-Pitaevskii equation \cite{lepoutre} confirm. Under these assumptions, the BEC is fully characterized by the (spatially-dependent) orientation of  $\vec{S}$, the local collective spin normalized by the density, and the spin value ($|\vec{S}|=1$). The hydrodynamic equation, neglecting the corrections associated with dipole-dipole interactions (see below), reads \cite{kawaguchi2010}:
\begin{equation}
\frac{\partial \vec{S}}{\partial t}=-\vec{S} \times \left[ - \frac{\hbar}{2 M} \left( \vec{a}. \vec{\nabla} \right) \vec{S} - \frac{\hbar}{2 M} \nabla^2 \vec{S} + \frac{g \mu_B}{\hbar} \vec{B}(\vec{r})   \right]
\label{eqnhydro}
\end{equation}
with $\vec{a} = \vec{\nabla}(n_{tot})/n_{tot}$; $g$ is the Land\'e factor and $\mu_B$ the Bohr magneton. $\vec{B}(\vec{r})$ captures magnetic field inhomogeneities. We do not include the spatially homogeneous magnetic field, which simply introduces homogeneous Larmor precession and can be gauged out. We point out that while Eq. (\ref{eqnhydro}) does not depend on the interaction parameters, the hydrodynamic behavior is inherently tied to strength of the mean-field interparticle interactions.

Initially, all spins are tilted perpendicular to the $x$ axis. In order to find the collective oscillatory modes associated to Eq. (\ref{eqnhydro}), we make the following ansatz: $\vec{S} (\vec{r}) = \left\{ f , g ,  \sqrt{ 1-f^2-g^2 }  \right\} $ with $f \equiv P(\vec r) \sin \omega t$ and $ g \equiv P(\vec r) \cos \omega t$. To obtain analytical solutions,
we make in addition a Gaussian ansatz for the atomic distribution: $n_{tot}=N/\pi^{3/2}/(\sigma_x \sigma_y \sigma_z) \exp(-x^2/\sigma_x^2-y^2/\sigma_y^2-z^2/\sigma_z^2)$. Minimizing the total energy one finds, in the limit
of large number of particles $N$, $\sigma_i\simeq R_i/\sqrt{3.23}$ where $R_i$ is the Thomas-Fermi radius.
In order to find the modes we assume vanishing magnetic field inhomogeneities, $\vec{B}(\vec{r})=0$ and
$f,g\ll S$. We obtain the following equation:

\begin{equation}
\label{poly}
\frac{M\omega}{\hbar}P(\vec r)=\left(\frac{x}{\sigma_x^2}\frac{\partial}{\partial x}+\frac{y}{\sigma_y^2}\frac{\partial}{\partial y}+\frac{z}{\sigma_z^2}\frac{\partial}{\partial z}- \frac{1}{2}\nabla^2\right)P(\vec r)
\end{equation}

Solutions of Eq. (\ref{poly}) are the well known Hermite polynomials and values for the frequencies are:

\begin{equation}
2 \pi \nu_{i,j,k} = \frac{\hbar}{M} \left( \frac{i}{\sigma_x^2} + \frac{j}{\sigma_y^2} + \frac{k}{\sigma_z^2}\right)
\label{freq}
\end{equation}
where the highest term of the polynomial is of the form $x^iy^jz^k$.
We point out that $\hbar^2/M \sigma^2$ is the energy scale associated with the confinement of the BEC wave-function. In the Thomas Fermi (TF) regime, $ 2 \pi \nu \approx \omega \times \frac{\hbar \omega}{\mu} \ll \omega $ (where $\mu$ is the chemical potential), which is a unique collective feature of spinor gases (in contrast to scalar BECs in their ground state). We also point out that, contrarily to \cite{marti2014} where a magnon wavevector is imprinted by Bragg pulses, here the wavevector ($ \propto 1/R$) of the trapped magnon spontaneously emerges from the BEC size.

The Gaussian ansatz is useful to make analytical predictions to compare to the experimental data. We numerically
calculated the three lowest mode frequencies by means of Bogoliubov analysis and find that they differ by less that 3 percent compared to Eq. (\ref{freq}) for our experimental parameters.

We now calculate the dynamics after rotation of the spins, in presence of a magnetic field gradient in the $x$ direction $\vec{B}(\vec{r})= b x \vec{u}_x$. We find that $f(x,t)= M \sigma_x^2 g \mu_B b/\hbar^2 \left(1-\cos 2 \pi \nu t\right)x$ and $g(x,t)=  \frac{M \sigma_x^2}{\hbar} \partial f(x,t)/\partial t $ constitutes the solution of Eq. (\ref{eqnhydro}), with $2 \pi \nu = \frac{\hbar}{M \sigma_x^2}$. The condition for the amplitude of the oscillations to be small is that $g \mu_B b \sigma_x \ll \hbar^2 /M \sigma_x^2$. In this regime, we have thus shown that the magnetic field gradient triggers the first ferrofluid collective mode characterized by the smallest frequency given by Eq. (\ref{freq}).

We also calculate the population dynamics. We use the fact that the local rotation of a spinor $\Psi$ in a stretched (fully polarized) state by an angle $\theta (x) \approx f_x x$ leads to a well defined modification of the local populations in the different Zeeman states along the quantization axis $p_{m_s} (\theta (x))$. We integrate such population changes as a function of space $\int p_{m_s} (\theta (x)) n_{tot}(\vec r) d^3r$ and we find an analytical expression for the dynamics of $p_{m_s}(t)$  \cite{suppmat}, which goes beyond the short time expression we obtained in \cite{lepoutre}. Similarly, we find that the density profile of each Zeeman component $m_s$ oscillates in time, with a peak density at a position $x_{m_s}(t) \approx u(t) \sigma_x m_S/(1+3 u(t)^2)$, with $u(t) =  \left(\frac{g \mu_B b M \sigma_x^3}{\hbar^2}\right) \left( 1 - \cos 2 \pi \nu t \right)$.  This motion, linked to the time evolution of $\delta(t)$, is sinusoidal only for $b \rightarrow 0$.

Finally, to explore how dipole-dipole interactions (which often cannot be neglected in chromium \cite{chromium1,chromium2,chromium3,chromium4}) modify the spin collective modes, we have studied the Bogoliubov excitations of a ferromagnetic BEC for the homogeneous case. In absence of dipole-dipole interactions, there are two well-known un-gapped modes, one corresponding to density excitations and one corresponding to magnons, with $\epsilon(k)= \hbar ^2 k^2/2 M$ \cite{kawaguchi2012}. In presence of dipole-dipole interactions, we find that the magnon has the following modified dispersion relation:
\begin{equation}
E(k)=\sqrt{\left(\epsilon(k)- n S c_{dd} \pi F(\theta) \right)\left(\epsilon(k) +2 n S c_{dd} \pi F(\theta)\right)}
\end{equation}
where $n$ is the density, $c_{dd} = \mu_0 (g \mu_B)^2 / (4 \pi)$ the dipolar interaction intensity, and $\theta$ the angle between the momentum and the magnetic field. $\mu_0$ is the magnetic constant. $F(\theta)=(1/3 + \cos 2 \theta)$. As a consequence, the magnon mode becomes dynamically unstable ($i.e.$ $E(k)$ becomes imaginary) at low momentum $k< k_m$. The characteristic momentum associated to such instability corresponds in our experimental situation to a distance $2 \pi /k_m \approx 6.5$ $\mu$m which is slightly larger than the size of the sample. Thus, our trap tends to protect the mode from being unstable. However, for large enough samples, dipole-dipole interactions can strongly modify the nature of the collective excitations.

We now turn to the comparison between our theoretical model and our experimental observations. As chromium atoms undergo non-negligible atomic losses due to dipolar relaxation \cite{lepoutre, pasquiou2010}, the density profile of the cloud is time-dependent. We use Eq.(\ref{freq}) with $\sigma$ corresponding to the direction where the magnetic field gradient $b$ is strongest and evaluated for the average atom number during dynamics. We expect a better agreement at the smallest $b$ for two reasons: (i) the linear regime is more likely to be valid; (ii) the protection of ferromagnetism \cite{lepoutre} is then most effective.

We show in Fig. \ref{gradient} that the observed frequency when $b \rightarrow 0$ matches well with the value obtained from Eq.
(\ref{freq}), 36 $\pm 6$ Hz (the uncertainty is to take into account the time-dependence of $R_{TF}$). We also point out the increase of the observed frequency as a function of $b$. This is in agreement with the expected frequency shift which arises because a larger $b$ leads to a larger mode amplitude \cite{suppmat}. At the largest $b$, the amplitude is large and the frequency approaches the trap frequency. In this regime, our ferrofluid model is most likely not valid, and the motion approaches that of free particles in an harmonic trap. In the intermediate regime, our data unveil beating between different frequencies (see Fig \ref{principe}), which indicates that two modes or more are simultaneously excited.

\begin{figure}[t]
\centering
\includegraphics[width= 8.5 cm]{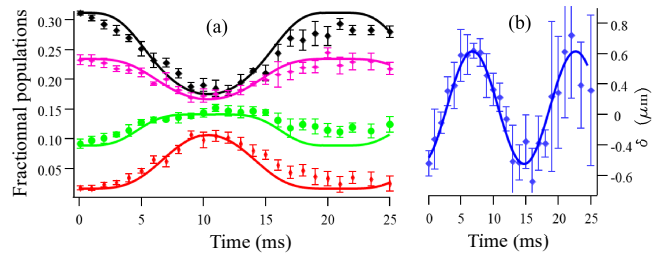}
\caption{Spin dynamics for a magnetic field gradient of $g \mu_B b/h=$ 5.5 MHz m$^{-1}$. (a) Fractional spin populations; from top to bottom $p_0$, $p_{-1}$, $p_{-2}$, $p_{-3}$, the solid lines result from the analytical expression detailed in \cite{suppmat}. (b) Spatial spin separation $\delta(t)$ is shown, together with a sinusoidal fit. In both cases, error bar correspond to statistical fluctuations.}
\label{agreement}
\end{figure}

We demonstrate in Fig. \ref{agreement}(a) that the observed experimental population dynamics is in very good agreement with the analytical expression in \cite{suppmat}. To perform such comparison, we used Eq. (\ref{freq}) for the value of $\nu$, and an adjusted value of $\sigma= 2.45$ $\mu$m for $u(t)$, which compares well to the value  obtained by dividing the geometrical average of the TF radii by $\sqrt{3.23}$ (see above), $2.3$ $\mu$m. At this magnetic field gradient, the amplitude of oscillations of $\delta (t)$, about 1.1 $\mu$m (see Fig. \ref{agreement}(b)), is also in good agreement with the simple estimate given by the amplitude of $\frac{d (x_{m_s+1}-x_{m_s})}{dt} \times t_{tof}$, where $x_{m_s}(t)$ has been estimated above: using $\sigma = 2.3$ $\mu$m, we find an amplitude of oscillation of 0.9 $\mu$m.

Taken together, the agreement between our observations and our model provides compelling evidence that a ferrofluid mode, coupling the spin and orbital degrees of freedom, is excited by  the magnetic field gradient.  This collective mode driven by mean-field interactions is well understood in the framework of superfluid hydrodynamics. This demonstrates that the hydrodynamic approach can provide a very powerful description of multi-component spinor gases. In the case of an initial ferromagnetic spin texture, it also provides a unified description of spinor gases, which is, under certain approximations, independent of the nature of interactions between atoms. For example, the ferrofluid trapped magnon mode is first seen in our experiment, which uses a spinor gas with relatively strong dipole-dipole interactions together with strong spin-dependent contact interactions which energetically \textit{disfavor} ferromagnetic behavior. However, we expect that this mode can be observed with any spinor gas, initially in a stretched (ferromagnetic) state, and independent of the sign of spin-dependent short-range interactions \cite{stamperkurn,kawaguchi2012}. We also point out that the presence of dipole-dipole interactions can destabilize the observed trapped magnon mode, which could be studied using a trap of larger volume, or atoms with larger magnetic moments such as dysprosium \cite{dysprosium} or erbium \cite{erbium}.

Acknowledgements: We thank A. M. Rey and B. Zhu for stimulating discussions. We thank R. Dubessy and M. Robert-de-Saint-Vincent for a critical reading of our manuscript. We acknowledge financial support from Conseil R\'egional d'Ile-de-France under DIM Nano-K/IFRAF, CNRS, Minist\`ere de l'Enseignement Sup\'erieur et de la Recherche within CPER Contract, Universit\'e Sorbonne Paris Cit\'e, and the Indo-French Centre for the Promotion of Advanced Research under LORIC5404-1 and PPKC contracts.

\section{Supplemental material}

\subsection{Extension to large amplitudes}

To obtain the mode frequencies beyond the linear regime,
we still assume a Gaussian ansatz for the density but we do not linearize anymore Eq. (\ref{eqnhydro}). While solution are not anymore polynomials we make truncated series expansion in spatial coordinates, i.e. we assume polynomial solutions for the functions $f$ and $g$ (see above). Furthermore, within the Gaussian ansatz, Eq. (\ref{eqnhydro}) only couples terms $x^{p}$ with $x^{p \pm 2}$, so that we can chose polynomials including either only odd or only even powers of $x$.

Here, as an illustrative example, we study solutions with a polynomial of order 3 for the spin
$\vec{S}  = \left\{ f , g , \sqrt{1-f^2-g^2}  \right\} $:
\begin{eqnarray}
f= (\alpha x+ \gamma x^3) \sin \omega t \nonumber \\
g = (\alpha x + \gamma x^3) \cos \omega t \nonumber
\end{eqnarray}
We thus find:
\begin{eqnarray}
\omega \alpha = - \frac{\hbar}{2 M R^2} (-2 \alpha + \alpha^3 \sigma^2+ 6 \gamma \sigma^2) &          & \nonumber \\
\omega \gamma = - \frac{\hbar}{4 M R^2} (-2 \alpha^3 -12 \gamma + 3 \alpha^5 \sigma^2 + 20 \alpha^2 \gamma \sigma^2) \nonumber
\end{eqnarray}
Keeping the lowest orders, we find:
\begin{eqnarray}
\label{beyondsmall}
2 \pi \nu_1 = \frac{\hbar}{M \sigma^2} \left( 1 + \alpha^2 \sigma^2/4 \right)   \\
2 \pi \nu_3 = 3 \frac{\hbar}{M \sigma^2}  \left( 1 -23/12 \alpha^2 \sigma^2 \right) \nonumber
\end{eqnarray}

The coefficient $\alpha$ is proportional to the magnetic gradient $b$: $\alpha=M \sigma^2g\mu_B b/\hbar^2$ (see above).
When $b$ increases the frequency of the lowest mode increases as well.
 We show in the main paper
the result obtained for Eq. (\ref{beyondsmall}) (see Fig 3) with $g mu_B b/h=$ 2.8 MHz/m for $b=$ 1 Gauss/m.

\subsection{Analytical equation for population dynamics}

The hydrodynamic equation and the Gaussian ansatz for the density allow calculating analytically the population dynamics after the initial rotation of the spins. Following the discussion in the main paper, we calculate   $p_{m_s} (t) = \int p_{m_s} (\theta (x)))  n(\vec r) d^3r $ with $\theta (\vec x) =  xM \sigma^2 b/\hbar \left(1-\cos 2 \pi \nu t\right)$. Here, $p_{m_s} (\theta (x))$ is the local $m_s$ population along the quantization axis corresponding to a local rotation of the spins by $\theta (x)$. We find:

\begin{eqnarray}
p_0 = \frac{5}{512} e^{-9 u^2} \left[1+ 6 e^{5u^2}+15 e^{8 u^2} + 10 e^{9 u^2} \right] \nonumber \\
p_{1,-1} = \frac{15}{2048} e^{-9 u^2} \left[-1+ 2 e^{5u^2}+17 e^{8 u^2} + 14 e^{9 u^2} \right] \nonumber \\
p_{2,-2} = \frac{3}{1024} e^{-9 u^2} \left[1-26 e^{5u^2}+15 e^{8 u^2} + 42 e^{9 u^2} \right] \nonumber \\
p_{3,-3} = \frac{1}{2048} e^{-9 u^2} \left[-1+ 66 e^{5u^2}-495 e^{8 u^2} + 462 e^{9 u^2} \right] \nonumber
\label{population}
\end{eqnarray}
with $u(t) =  \left(\frac{g \mu_B b M \sigma^3}{\hbar^2}\right) \left( 1 - \cos 2 \pi \nu t \right)$. We emphasize that the population dynamics, although inherently set by spin-dependent interactions, display an expression where the interaction parameter are absent, one striking consequence of the ferro-hydrodynamic equations.

\end{document}